\documentclass[reprint,prl,longbibliography,superscriptaddress]{revtex4-2}

\usepackage{amssymb, amsmath}
\usepackage[english]{babel}
\usepackage{bm}
\usepackage{graphicx}
\usepackage[utf8]{inputenc}
\usepackage{bbold}

\usepackage[dvipsnames]{xcolor}

\newcommand{\commenttoggle}[1]{}

\newcommand{\red}[1]{{ #1}}
\newcommand{\redblock}{}

\mathchardef\myminus="2D

\usepackage{letltxmacro}
\LetLtxMacro{\ORIGselectlanguage}{\selectlanguage}
\makeatletter
\DeclareRobustCommand{\selectlanguage}[1]{%
  \@ifundefined{alias@\string#1}
    {\ORIGselectlanguage{#1}}
    {\begingroup\edef\x{\endgroup
       \noexpand\ORIGselectlanguage{\@nameuse{alias@#1}}}\x}%
}
\newcommand{\definelanguagealias}[2]{%
  \@namedef{alias@#1}{#2}%
}

\makeatother

\definelanguagealias{en}{english}
\definelanguagealias{EN}{english}
\definelanguagealias{eng}{english}
\definelanguagealias{de}{ngerman}

\begin{document}
\author{Marten Richter}
\email[]{marten.richter@tu-berlin.de}

\affiliation{Institut für Theoretische Physik, Nichtlineare Optik und
Quantenelektronik, Technische Universität Berlin, Hardenbergstr. 36, EW 7-1, 10623
Berlin, Germany}

  \author{Stephen Hughes}
 \affiliation{Department of Physics, Engineering Physics, and Astronomy, Queen's University, Kingston, Ontario K7L 3N6, Canada}

\title{Enhanced TEMPO algorithm for quantum path integrals with off-diagonal system-bath coupling: applications to photonic quantum networks}


\begin{abstract}
Multitime system correlations functions  are relevant in various areas
of physics and science, dealing with system-bath interaction including spectroscopy and quantum optics, where many of these schemes include an
off-diagonal system bath interaction. 
Here we extend the enhanced TEMPO algorithm for quantum path integrals using tensor networks [Phys. Rev. Lett. 123, 240602 (2019)] to open quantum systems  with 
off-diagonal coupling 
beyond a single two level system.
We exemplify the approach
on a coupled cavity waveguide system with spatially separated quantum two-state emitters, though many  other applications in material science are possible, including entangled photon propagation, photosynthesis spectroscopy and on-chip quantum optics with realistic dissipation.

 \end{abstract}


\date{\today}
\maketitle

The theory of open quantum system
remains a very active 
focus of current research \cite{breuer2002theory,CALDEIRA1983587,tanimura1993real,RICHTER2010711,makri1995tensor,makri1995tensor2,vagov2011real,strathearn2017efficient,strathearn2018TEMPO,kaestle2020protected,PhysRevB.102.235303}, since it provides answers to many questions related to 
quantum computing, entanglement and communication, and also  quantum networks. The 
theoretical frameworks 
include tensor network (TN) methods \cite{orus2014practical,schollwock2011density,CIRAC2017100,verstraete2006matrix,vidal2007classical,CIRAC2017100,plenioheisenberg,PhysRevLett.116.237201,PhysRevLett.116.093601,kaestle2020protected,schroder2019tensor} and quantum path integrals \cite{CALDEIRA1983587,tanimura1993real,makri1995tensor,makri1995tensor2,vagov2011real,strathearn2017efficient,strathearn2018TEMPO,kaestle2020protected,PhysRevB.102.235303}. Recently these two approaches were successfully combined \cite{strathearn2018TEMPO,PhysRevLett.123.240602ETEMPO,PhysRevB.102.235303,kaestle2020protected}, exploiting their combined advantages in the TEMPO (time-evolving matrix product operator) algorithm \cite{strathearn2018TEMPO,strathearn2020modelling} and an enhanced TEMPO (eTEMPO) algorithm \cite{PhysRevLett.123.240602ETEMPO}. Applications so far have mainly studied a two level system (TLS) {\it diagonally} coupled  to an external bath \cite{strathearn2017efficient,strathearn2018TEMPO,strathearn2020modelling,PhysRevLett.123.240602ETEMPO,kaestle2020protected,PhysRevB.102.235303,juliainductive}.
Beside these examples, many important problems remain to be studied in more detail, ranging from exciton relaxation in photosynthetic light harvesting systems \cite{Chernyak:1996,zhang1998exciton,panitchayangkoon2010long,Kramer2018} or nanostructures to exploit photon propagation, entanglement,
superradiance and feedback \cite{oulton2008hybrid,stockman2004nanofocusing,orieux2017semiconductor,weiss2018interfacing,jayakumar2014time,CarmeleReitzenstein,PhysRevA.83.063805,Gegg_2018,PhysRevLett.116.093601,kaestle2020protected,PhysRevResearch.3.023030}---which require {\it off-diagonal} system bath coupling. Mostly their treatment requires other algorithms such as hierarchical equations of motion \cite{tanimuraheom,Kramer2018}, and alternative TNs \cite{PhysRevLett.116.093601,kaestle2020protected}. 

Here we extend the eTEMPO method to include off-diagonal coupling \red{as opposed to extending the TEMPO algorithm \cite{gribben2021exact}}. As input, we require  only the generalized bath correlation function; 
in contrast to Ref.~\onlinecite{cygorek2021numerically}, where   bath 
degrees of freedom are included \red{and accessible} in the network propagation,
we show the effect of retardation for a
two-cavity waveguide system on the first 
two rungs of photon transition. We demonstrate how the system transitions between one single generalized Jaynes-Cummings model (JCM)  to a retardation regime 
between two JCMs, including a subradiant state. This manifests in a  highly non-trivial non-Markovian dynamic, whose
features {\it cannot} be captured with linearized response functions nor phenomenological JCMs.
\begin{figure}[bth]
	\includegraphics[width=8.5cm]{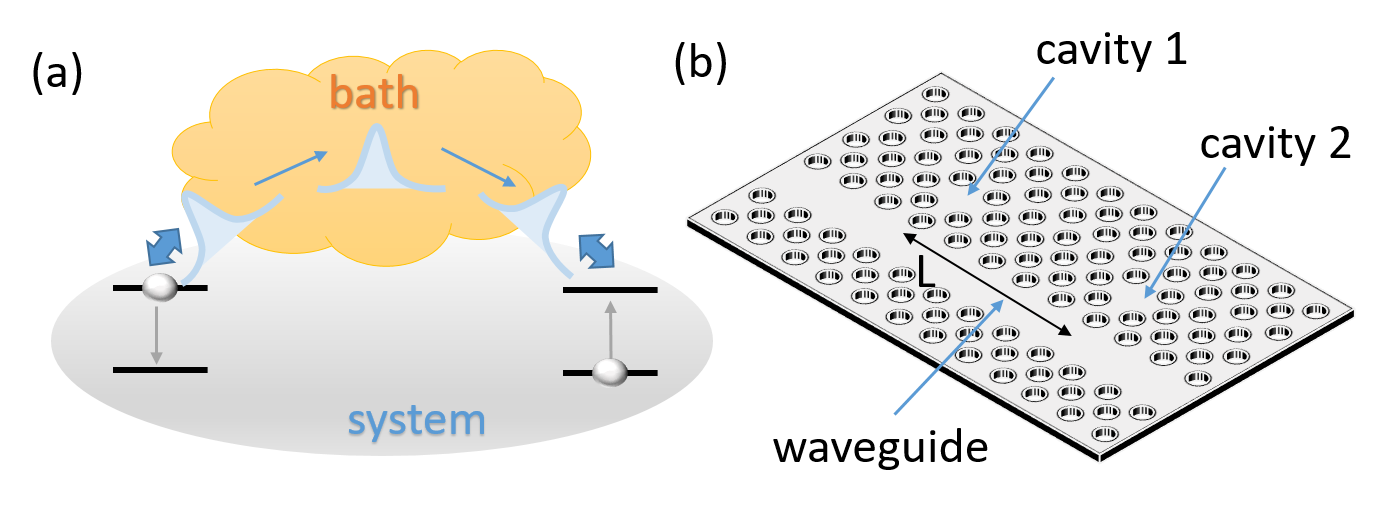}
	\caption{(a) Open quantum system coupled to a bath with excitation transfer through the bath. (b) Example system: photonic crystal with waveguide and two cavities separated by length $L$ with a quantum emitter (TLS) in each cavity. }	\label{schemesystem}
\end{figure}

Theoretically, a typical open quantum system, has system
$H_{s}$  and bath $H_{b}$ Hamiltonian.
\red{Figure \ref{schemesystem} shows an
example integrated waveguide system,
which is representative of emerging
experiments with integrated semiconductor
quantum dot systems~\cite{Fox2018,Dietrich2016, PhysRevLett.108.227402, PhysRevB.97.235448, Kim2016, Sato2011, Schall2021, Vora2015, Kim2018, Khoshnegar2017, PhysRevB.99.085311}.}
The bath is harmonic, so Wick's theorem holds for factorizing initial conditions \cite{PhysRevE.68.021111}.
We consider a linear system bath interaction $H_{sb}=\sum_{ij\mu} C_{ij\mu} A_{ij} B_{\mu}$, where  $C_{ij\mu}$ are the  system-bath coupling constants,
$A_{ij}=|i\rangle_s\langle j |_s$  (here in particular including $i\neq j$) and $B_{\mu}$ is a linear bath operator; for an harmonic bath $\mathrm{tr} (B_{\mu}\rho_B)=0$.  
The dynamics of the full system-bath density matrix operator $\rho$ obeys $\partial_t \rho=-\frac{i}{\hbar} (H_{s,-}+H_{b,-}+H_{sb,-})\rho$,
where the subscripts $-,+, L, R$ convert a Hilbert space operator $D$ to a Liouville space operator \cite{Chernyak:1996} with $D_L\rho =D\rho$ and $D_R\rho=\rho D$, and also $D_-=D_L-D_R$. Notably, the method can also include a Lindblad operators for system dynamics together with $H_{s,-}$.
The time dynamics can be solved via the time ordered exponential $U(t,t_0)=T_\leftarrow \mathrm{exp}\left(-\frac{i}{\hbar}\int_{t_0}^t H_-(\tau) \mathrm{d} \tau \right)$  in Liouville space, $\rho(t)=U(t,t_0)\rho(t_0)$.
Our observables of interest are multitime correlation functions $\langle T_\leftarrow A_1(t_1)\cdots A_N(t_N) \rangle$ with Liouville operators $A_1$, ...,$A_N$.

Next, we convert the correlation function into a quantum path integral formulation \cite{CALDEIRA1983587,tanimura1993real,makri1995tensor,makri1995tensor2,vagov2011real,strathearn2017efficient,strathearn2018TEMPO,kaestle2020protected,PhysRevB.102.235303}; this allows us to develop an extended algorithm for systems with off-diagonal system-bath coupling, based on the eTEMPO algorithm \cite{PhysRevLett.123.240602ETEMPO}---which is a very efficient algorithm for many quantum path integral TN implementations~\cite{strathearn2018TEMPO,kaestle2020protected,PhysRevB.102.235303}.
To proceed, we divide time $t$ into intervals $\Delta T$ such that $t_n=n\cdot \Delta T$ with integer $n$. The times in the multitime correlation function should obey $t_i=\Delta T\cdot n_i$ with integer $n_i$, and we obtain 
$\langle T_\leftarrow A_1(t_1)... A_N(t_N) \rangle=\mathrm{tr}(O_{M} U_{M\myminus1} ...  U_1O_1U_0\red{O_0}\rho(t_0)), 
$
with $O_n=A_k$ if $t_k=\Delta T\cdot n$ for any $k$, otherwise $O_n=Id$;
$U_i$ is defined as $U_i=U(t_{i+1},t_{i})$.
For most path integral implementations, the Suzuki-Trotter formula is applied to separate system and bath for the influence functional.
We use perturbation theory and the Feynman disentanglement theorem. Perturbation theory to a limited order 
fails for processes involving many interactions over all times; thus, 
 we apply perturbation theory to the individual intervals $\Delta T$, so to $U_i$, keeping often only one system bath process per $\Delta T$.   For sufficient small $\Delta T$, compared to system-bath coupling, a nonperturbative result is obtained with numerical accuracy.
In general,
$U_i=U_0(t_{i+1},t_{i+\frac{1}{2}})~ U_I^{(i)}~  U_0(t_{i+\frac{1}{2}},t_{i})$,
with 
$
U_I^{(i)}{=}T_\leftarrow \mathrm{exp} \left( -\frac{i}{\hbar} \int_{t_i}^{t_{i+1}}\! U_0(t_{i+\frac{1}{2}},\tau) H_{sb,-} U_0(\tau,t_{i+\frac{1}{2}})  \mathrm{d} \tau\right). 
$

Restricting the system-bath coupling $H_{sb,-}$ to first order, per $\Delta T$ (and second order for non-vanishing bath correlation function, as  explained \red{after Eq. \eqref{influ_ind}}), yields:
\red{
\begin{align}
&U_i=U_0(t_{i+1},t_{i+\frac{1}{2}})U_{sb,i}
 U_0(t_{i+\frac{1}{2}},t_{i})\nonumber\\
&U_{sb,i}=\left( Id-\frac{i}{\hbar}\int_{t_i}^{t_{i+1}}   \mathrm{d}\tau U_0(t_{i+\frac{1}{2}},\tau) H_{sb,-} U_0(\tau,t_{i+\frac{1}{2}})
\right.\nonumber\\
&\quad\quad -\frac{1}{\hbar^2} \int_{t_i}^{t_{i+1}}   \mathrm{d}\tau_1 \int_{t_i}^{\tau_1}   \mathrm{d}\tau_2 U_0(t_{i+\frac{1}{2}},\tau_1) H_{sb,-} U_0(\tau_1,\tau_2) \nonumber\\
&\qquad\qquad \left.   H_{sb,-} U_0(\tau_2,t_{i+\frac{1}{2}})\right), \label{uexpansion}
\end{align}
with the time evolution operator $U_0(\cdot,\cdot)$ containing solely $H_{s,-}$ and $H_{b,-}$. The symmetric expansion with $U_0(\cdot,\cdot)$ on the left and right allows to exclude (include) certain parts of the Hamiltonian (e.g. like external optical excitation) in the inner brackets.
Including second-order contributions from the same $\Delta T$ (cf.~cumulant expansions~\cite{mukamel1995principles}) prevents an artificial minimum delay between system bath interactions and reducing $\Delta T$ dependency, and thus recovers simple perturbation theory for weak coupling.
We define $U^{\frac12}_{0,i+\frac12}=U_{0}(t_{i+\frac{1}{2}},t_i)$ and rewrite  $U_{sb,i}=U_{sb,i}^{(0)}+U_{sb,i}^{(1)}+U_{sb,i}^{(2)}$, with  the $k$-th order system-bath contribution  $U_{sb,i}^{(k)}$;
 $H_{sb,-}$ contains system $A$ and bath operators $B$, so we apply $(AB)_-=A_+B_-+A_-B_+$.  Finally, $U_{sb,i}^{(k)}$ can be written as  $U_{sb,i}^{(k)}=\sum_l A^{(k)}_{l} B^{(k)}_{l}$ with  Liouville system operators $A^{(k)}_{l}$ and bath operators  $B^{(k)}_{l}$(each with a maximum $k$ linear bath operators).
}

With an initial factorizing density matrix $\rho(t_0)=\rho_s\otimes\rho_b$, then 
{\redblock
\begin{align}
&\langle T_\leftarrow A_1(t_1)\cdots A_N(t_N) \rangle=\label{sys_prop_influ}
\\ &\ \sum_{k_{M-1}.\dots k_0 =0}^{2}\sum_{l_{M-1}\dots l_{0}}\mathrm{tr}_b(U_{b,M-\frac{1}{2}} B^{(k_{M-1})}_{l_{M-1}}\dots 
 U_{b,\frac{1}{2}} B^{(k_{0})}_{l_0}\rho_B  )\nonumber\\
&\ \cdot 
\mathrm{tr}_s(O_{M} U_{s,M-\frac{1}{2}} A^{(k_{M-1})}_{l_{M-1}} \dots 
 O_1U_{s,\frac{1}{2}} A^{(k_0)}_{l_{0}} U_{s,0}^{\frac12}O_0\rho_S  ),\nonumber 
\end{align}
}
holds with the bath $U_{b,i}$ and system  $U_{s,i}$ time propagation. \red{
For initial thermal correlated $\rho(t_0)$, a Liouville operator $A_c$ can be included in $O_0$ \cite{GRABERT1988115}, while keeping an harmonic $\rho_B$ in Eq. \eqref{sys_prop_influ}. }

\begin{figure}[tb]
	\includegraphics[width=8.2cm]{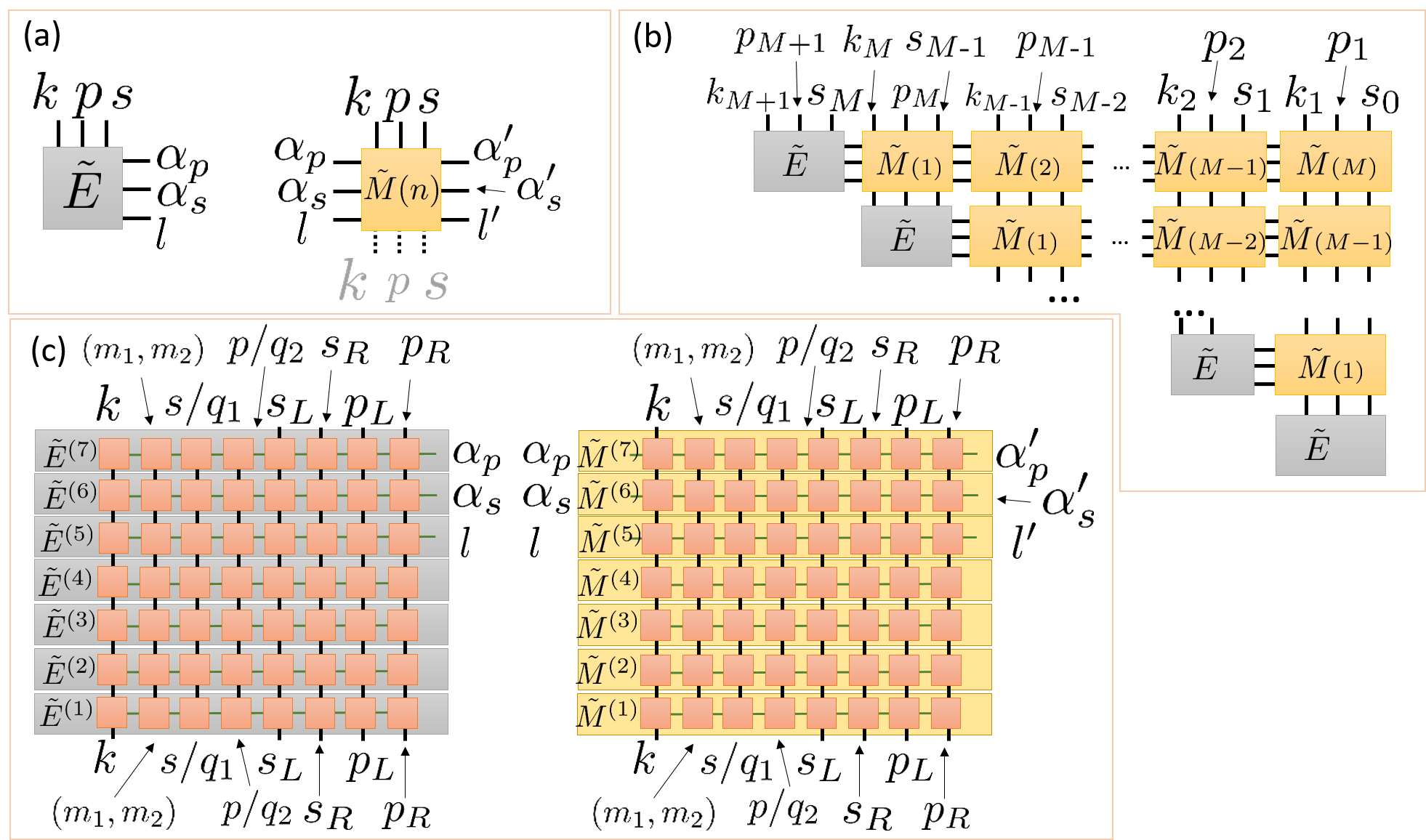}
	\caption{(a) Tensors $\tilde{E}$ and $\tilde{M}$	depicted as rectangle and indices. (b) TN for the influence functional from Eq. (\ref{influ_ind_ten}), c) decomposition of the $\tilde{E}$ and $\tilde{M}$ into seven matrix product operators (indices are connected with  delta tensors).}  
	\label{fig_influencefct1}
\end{figure}
\begin{figure}[tb]
	\includegraphics[width=7.5cm]{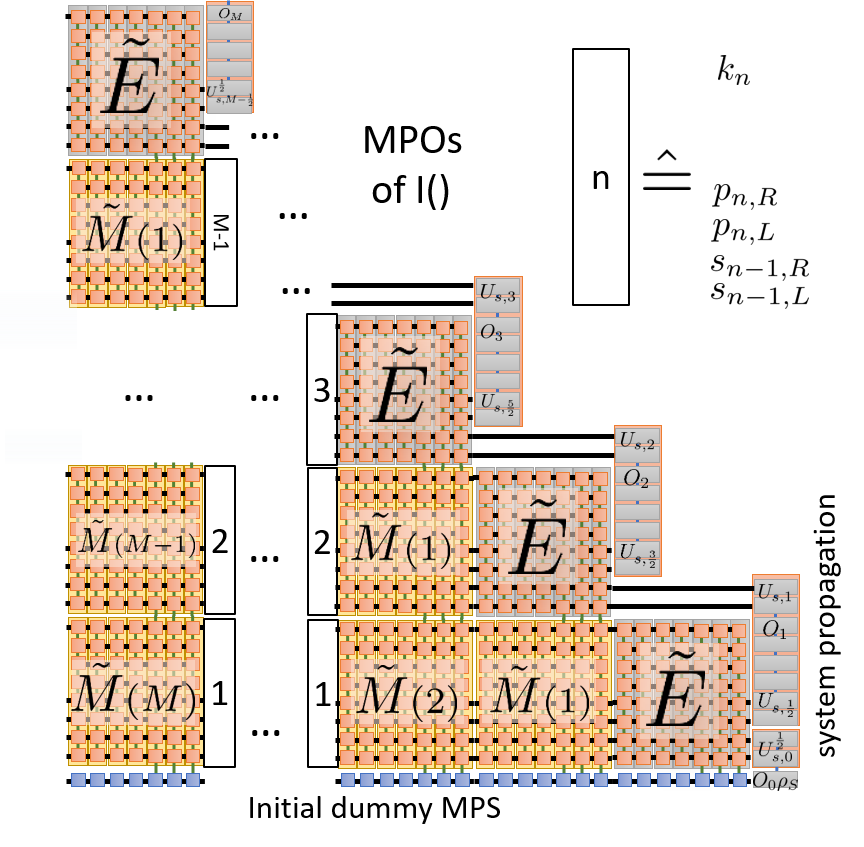}
	\caption{TN for $\langle T_\leftarrow A_1(t_1)... A_N(t_N) \rangle$. On the right edge of the TN the MPS decomposed system propagators $U_{s,m-1/2}$ (cf. Eq.(\ref{sys_prop_influ})) are connected to the indices $s$ and $p$. 
		The  majority of  network is constructed from the decomposed $\tilde{M}$ and $\tilde{E}$ tensors. An initial MPS is at the bottom.}
	\label{fig_influencefct2}
\end{figure}
To convert the expressions to a quantum path integral, we insert Liouville space identities: $Id=\sum_{s_L, s_R} (|s_L\rangle \langle s_L|)_L (|s_R\rangle \langle s_R|)_R$, using a $H_s$ eigenbasis.
Using the notation $A=\sum_{s,p} A^{p_Ls_L}_{p_Rs_R}  (|p_L\rangle \langle s_L|)_L (|s_R\rangle \langle p_R|)_R$ for expanding any operator $A$, we obtain
$
\langle T_\leftarrow A_1(t_1)\cdots A_N(t_N) \rangle=
\sum\limits_{\{k\}\{s\} \{p\} } I(s_0,p_1,s_1,...,p_M,s_M )
\mathrm{tr}_s\!\left[\tilde{O}_{s_M p_M}  ...  \!\tilde{O}_{s_1 p_1} \!\tilde{\rho}_{s_0p_0}  \!\right],\nonumber
$
 with the convention $\{x\}=x_1\dots x_M$ for indices and
with $\tilde{O}_{s_n p_n}=\left(O_{n} U_{s,n-1/2}\right)_{s_n p_n}$ and $\tilde{\rho}_{s_0p_0}=\left( U_{s,0}^{\frac12}O_0\rho_S \right)_{s_0p_0}$.
Without subscripts $L$ or $R$, an index includes left and right Liouville space.
We  replace the index $l$ and the operators $A^{(k)}_{l}$ with indices $s$ and $p$,  
and an according redefinition of operator matrix elements $B^{(p_js_j)}_{k_j}$ including the free bath propagation (and indicating the interval with index $j$)   in the influence functional $I$, so that
$
I(s_0,p_1,k_1,\dots,p_M,s_M, k_M )=
\mathrm{tr}_B( B^{(p_Ms_{M-1})}_{k_{M-1}}\dots  B^{(p_2s_1)}_{k_1}   B^{(p_1s_0)}_{k_0}\rho_b ).
$

Wick's theorem holds for factorization, since the harmonic bath is initially in thermal equilibrium (no photons).
Thus, $I$ factorizes into expectation values of two linear bath operators, with  each linear bath operator from a different interval $\Delta T$ or  two from the same interval $\Delta T$.
 We arrive at an iterative expression for $I$:
\begin{align}
&I(s_0,p_1,k_1)=E_{s_0p_1k_1},\label{influ_ind}\\
&I(s_0,p_1,k_1,...,s_M,p_{M+1},k_{M+1} )\nonumber\\
&\quad=E_{s_M,p_{M+1},k_{M+1}}I(s_0,p_1,k_1,...,s_{M-1},p_{M},k_{M} )\nonumber\\
&\quad\quad+\delta_{k_{M+1},1}\sum_{m=1}^M\delta_{k_m,1} M_{s_{M},p_{M+1},k_{M+1}}^{p_{m+1}s_{m}}
\nonumber\\
& \qquad \quad \cdot I(s_0,p_1,k_1,..., s_{m-1},p_m,-1,...,s_{M-1},p_{M},k_{M} ), \nonumber 
\end{align}
where $E_{s,p,k}$ describes the current time interval with zero ($\delta_{sp}$), two ($B^{(p s)}_{2}$) or one  system-bath interactions, \red{and $k=-1$ is added to link to previous times}:
$
E_{s,p,k}=(\delta_{sp}\delta_{k,0}+\delta_{k,2} \mathrm{tr}_b(B^{(p s)}_{2}\rho_B) +\delta_{k_1,-1})  
$ .
Here $M_{s_{M},p_{M+1},k_{M+1}}^{p_{m+1}s_{m}}$ describes a process, with one interaction in the interval $m$ (e.g., photon emission) and one in the interval $M+1$ (e.g., photon absorption):
$M_{s_{M},p_{M+1},k_{M+1}}^{p_{m+1}s_{m}}=\delta_{k_{M+1},1} \mathrm{tr}_b(B_{1}^{(s_{M},p_{M+1})} B_{1}^{(s_{m},p_{m+1})}\rho_b), $
and depends only on the time difference $M-m$ for time independent  bath Hamiltonians.

Note that $M_{s_{M},p_{M+1},k_{M+1}}^{p_{m+1}s_{m}}$ contains a generalized bath correlation function, directly connected to a generalized spectral density for off-diagonal coupling~\cite{Chernyak:1996},
which fully determines the system bath-interaction.
The tensor $M$ describes a boson going into the bath at $m$ and back to the system at $M+1$.
We discuss differences to the quantum path integrals with diagonal coupling---the standard influence functional \cite{CALDEIRA1983587,tanimura1993real,makri1995tensor,makri1995tensor2,vagov2011real,strathearn2017efficient,strathearn2018TEMPO,PhysRevLett.123.240602ETEMPO,kaestle2020protected,PhysRevB.102.235303}. For the common case, the $I$ depends only on one index $s$ per $\Delta T$, where for the off-diagonal case it depends on initial $s$ and final $p$ index  and the $k$ the number of system bath interactions per $\Delta T$. We note, hitherto, most numerically exact treatments with TN focused only on diagonal coupling.

We reformulate Eq.~\eqref{influ_ind} for easier conversion to a TN:
\begin{align}
&I(s_0,p_1,k_1,...,s_M,p_{M+1},k_{M+1} )=\nonumber\\
& \sum_{\{k'\} \{l\}\{\alpha_p\}\{\alpha_s\}} 
	\prod_{m=1}^M  \tilde{E}_{s_{M}p_{M+1}k_{M+1}}^{\alpha_{p_M}\alpha_{s_M}l}  \cdot
\tilde{M}_{p_{m+1}s_{m}k_{m}\alpha_{p_m}\alpha_{s_m}l_{m}}^{k'_m\alpha_{p_{m-1}}\alpha_{s_{m-1}}l_{m-1}}(n)\cdot
\nonumber\\
& \qquad   I(s_0,p_1,k'_1,...,s_{M-1},p_{M},k'_{M}  )\cdot \delta_{\alpha_{p_1}1}\delta_{\alpha_{s_0}1}\delta_{{l_1}1} , \label{influ_ind_ten}
\end{align}
with modified tensors $\tilde{M}$ and $n=M+1-m$:
\begin{align}
\tilde{M}_{p s k \alpha_{p} \alpha_{s}l}^{k' \alpha'_{p}\alpha'_{s}l'}(n)=&\delta_{k,1} \delta_{k',-1} \delta_{\alpha'_{p},1} \delta_{\alpha'_{s},1} \delta_{l_{m},1}
G_{ps l\alpha_{p}-1\alpha_{s}-1   }(n)\nonumber\\ &+\delta_{k,k'}\delta_{\alpha_{p}\alpha'_{p}} \delta_{\alpha_{s}\alpha'_{s}} \delta_{ll'},  \label{mtildetensor}
\end{align}
where $G(n)$ is the double integrated system bath correlation function (Eq.~(\ref{uexpansion})) between two $\Delta T$ intervals, which are $n\Delta T$ apart. The first interaction acts on left (right) side in Liouville space  for $l=1$ ($l=2$) respectively, changing the left/right state from $\alpha_{s}-1$ to $\alpha_{p}-1$. $\alpha'_{p/s}=1$ encodes no further interaction in subsequent $\Delta T$.   In Eq. \eqref{mtildetensor} the modified tensor $\tilde{E}$ appears:
\begin{align}
&\tilde{E}_{s,p,k}^{\alpha_{p},\alpha_{s}l }=\big [\delta_{k,0}\delta_{sp}+\delta_{k,2} G_{p s} +\delta_{k,-1} \label{etildetensor}
\\&\qquad+\delta_{k,1} (\delta_{l,1}\delta_{\alpha_p,p_L+1}\delta_{\alpha_s,s_L+1}+\delta_{l,2}\delta_{\alpha_p,p_R+1}\delta_{\alpha_s,s_R+1})  \big ]  \nonumber
\end{align}
where the $C$ tensor contains the system-bath correlation within the first interval $\Delta T$.

Equation~\eqref{influ_ind_ten} is now converted to a TN, where we depict a tensor as, e.g., $T_{ijk}$ as a rectangle and each index $i,j,k$  as a line (cf.~Fig.~\ref{fig_influencefct1}a)), where connected indices between tensors indicate a summation \cite{schollwock2011density}. The TN depicted in Fig.~\ref{fig_influencefct1}(b), built up from tensors $\tilde{E}$ and $\tilde{M}$, can be contracted by interpreting the first row  as a matrix product state (MPS) and the subsequent rows as matrix product operator (MPO), and applying MPOs to MPS subsequently ~\cite{schollwock2011density}. The TN is  the TEMPO \cite{strathearn2018TEMPO}  implementation for the off-diagonal case. 

\red{Because of the increased tensor rank in the off-diagonal case,
} the TEMPO algorithm is highly ineffective especially beyond a single TLS. Here
we construct a TN that reduces the index dimension and yields the original tensors after contraction.
%
%
Therefore, we design a product of 6 (and 7) low rank and low dimensional tensors  that yield $\tilde{M}^{(\cdot)}$ (and $\tilde{E}$). These  are also decomposed into a MPS~\cite{PhysRevLett.91.147902} and the resulting MPSs are connected via $\delta$-tensors to obtain the TN of MPOs in 
Fig.~\ref{fig_influencefct1}(c).
Further details 
are given in the Supplemental Information \cite{supplemental}.
%
\begin{figure}[tb]
	\includegraphics[width=8.3cm]{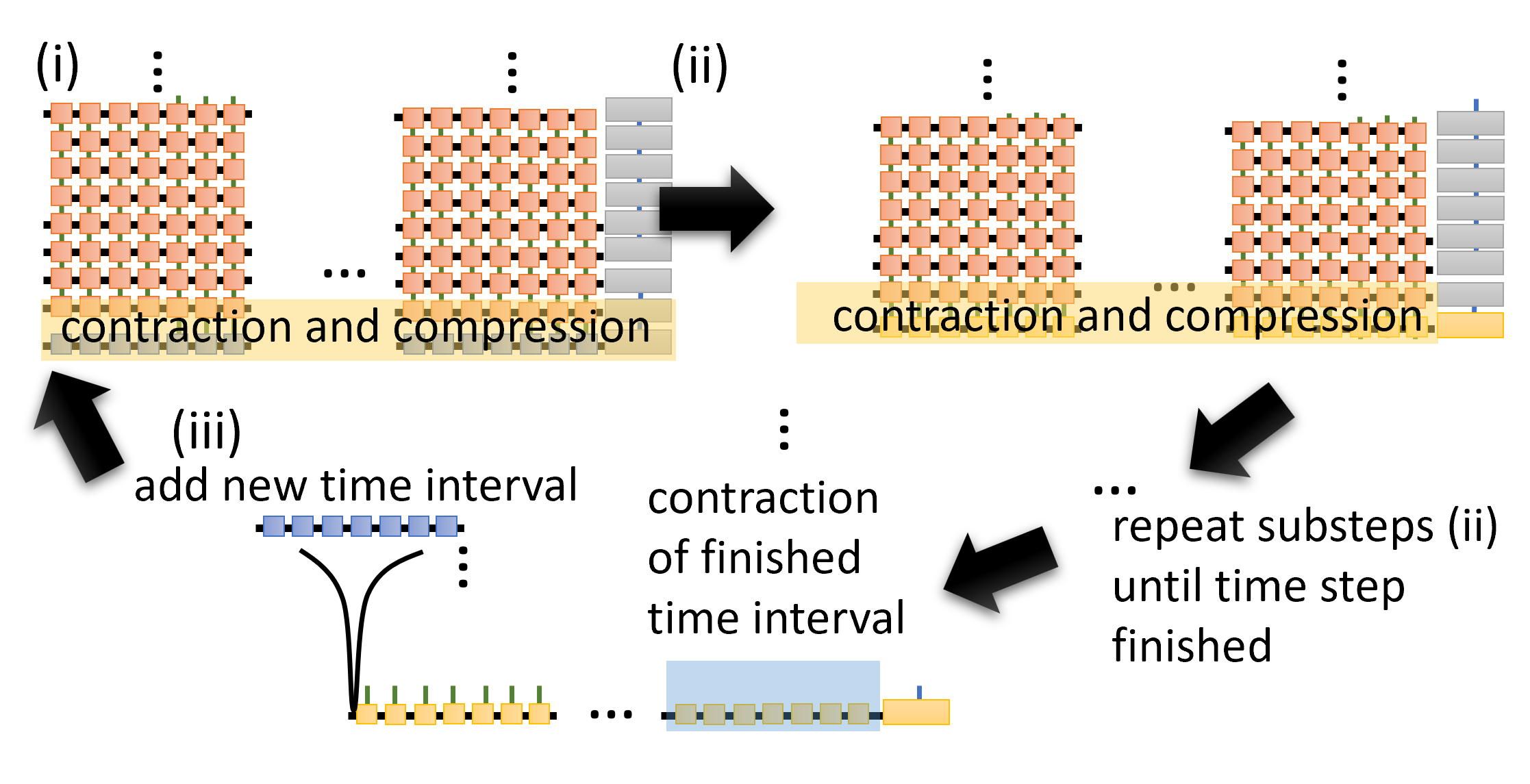}
	\caption{(i) Application of next MPO to the current MPS (initially a dummy MPS) (ii) repeated until all tensors of  $\Delta T$ are contracted, then (iii) removal  of the current interval, and extension for future intervals.  (i) Repeated, until finished.  }
	\label{fig_influencefct3}
\end{figure}
\begin{figure}[bt]
	\includegraphics[width=8.5cm]{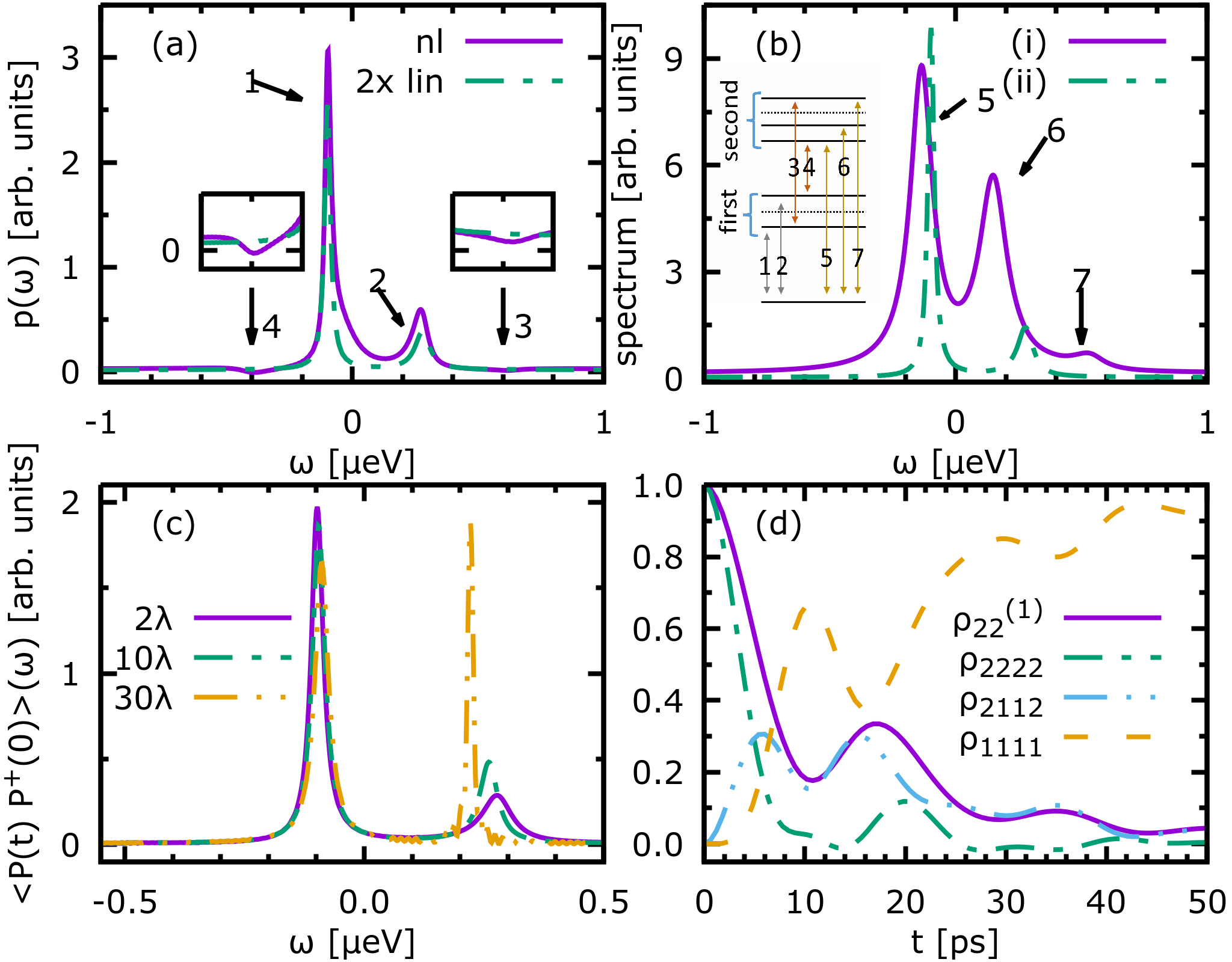}
	\caption{(a) Nonlinear (nl)  and linear (lin)  spectra for the setup from 
	Fig.~\ref{schemesystem}b) with a TLS resonant to both cavity frequencies. (b) Fourier transform of (i) double  
	$\langle P(t) P(t)P^{\dagger}(0) P^{\dagger}(0)\rangle$ and  (ii) single quantum  $\langle P(t) P^{\dagger}(0)\rangle$   emitter polarisation correlation function with $P^\dagger=\sum_{l} \mathbf{d} |2\rangle_l \langle 1|_l$ (inset JCM Ladder). (c) Single quantum  from (b) for different $L$. (d) Time dynamics of the upper level of reduced density matrix $\rho^{(1)}_{22}$ of one emitter, and  for both emitters $\rho_{nmmn}$  when both are initially excited. }
	\label{fig_qnmspec}
\end{figure}

Following the idea from Ref.~\onlinecite{PhysRevLett.123.240602ETEMPO}, that the indices on $\tilde{E}$ can be moved to the other edge of the network from 
Fig.~\ref{fig_influencefct1}(b) (rotating the TN by ninety degrees), we obtain the decomposed TN as in  Fig.~\ref{fig_influencefct2}. Then the TN (Fig.~\ref{fig_influencefct2}), including the temporal propagation of the system from Eq.~\eqref{influ_ind_ten}, is contracted to obtain the expectation values. 
 For evaluation, the MPOs are applied row by row (Fig.~\ref{fig_influencefct3} (i)-(iii),  Eq.~\eqref{influ_ind} using {\sl exactapply} from itensor ~\cite{itensor}),  thus implementing the eTEMPO algorithm \cite{PhysRevLett.123.240602ETEMPO} variant  for the offdiagonal case in itensor (version 2.1.0 patched)~\cite{itensor}. Tensors in the MPS connected to previous times are traced out, achieving a massive reduction of computational time (cf.~Fig. \ref{fig_influencefct3}).
Then a row of blocks in the network are added to the current MPS to extend the covered time, if required.

Critically, our framework is capable of including both diagonal and off-diagonal coupling and is thus applicable to a multitude of problems, including exciton-phonon dynamics in (coupled) nanostructures \cite{pssb.200668053,vagov2011real,kerfoot2014optophononics,zimmermann2016poisson,nazir2016modelling,luker2012influence}, photosynthetic pigment-protein complex \cite{zhang1998exciton,renger2002relation,panitchayangkoon2010long}, as well as quantum optics system including plasmonics \cite{oulton2008hybrid,stockman2004nanofocusing,orieux2017semiconductor,weiss2018interfacing,jayakumar2014time,PhysRevB.92.205420,kaestle2020protected,PhysRevLett.122.213901}.
To demonstrate the power of our approach, we 
consider {\it a quantum network example, with on-chip photonic propagation}, fully consistent with
a rigorous 
Maxwell solution
theory of a photonic crystal with a waveguide 
and two integrated cavities \cite{Yao:09} as depicted in Fig.~\ref{schemesystem}(b)  (system parameters in  \cite{supplemental}). \red{This scheme is 
also timely with 
recent experiments~\cite{Yu:21}.}  

\red{We assume a TLS with 
$H_s{=}\sum_{li }\epsilon_{i_l}|i\rangle_l \langle i|_l$, 
in each cavity, where $\epsilon_{i_l}$ is the energy of  level $i$ in system $l$.}
 \red{ The 
  photonic  crystal medium can be  quantized using the Green's function  $\mathbf{G}(\mathbf{r},\mathbf{r}',\omega)$ of the Helmholtz equation \cite{PhysRevA.57.3931,Yao:09,supplemental}.
The electric field operator is
$\mathbf{E}(\mathbf{r},\omega)=i\sqrt{\frac{\hbar}{\pi \epsilon_0}}\int \mathrm{d}^3r' \sqrt{\epsilon_I(\mathbf{r}',\omega)}\mathbf{G}(\mathbf{r},\mathbf{r}',\omega){\cdot}\mathbf{b}(\mathbf{r}',\omega)$, where $\mathbf{b}(\mathbf{r},\omega)$
are boson operators that
 form the bath for the path integral approach: $H_B{=}\hbar\int \mathrm{d}^3r\int^\infty_0 \mathrm{d}\omega ~ \omega \mathbf{b}^\dagger(\mathbf{r},\omega){\cdot} \mathbf{b}(\mathbf{r},\omega)$. 
 A dipole coupling $\mathbf{d}^{(i)}_{ij}$ for the $i$ emitter at position $\mathbf{r}_i$ to the bath constitutes the system-bath interaction $H_{sb}= \sum_{nij}  |i\rangle_n \langle j|_n \mathbf{d}^{(n)}_{ij}{\cdot}\int_0^\infty\mathrm{d}\omega \mathbf{E}(\mathbf{r}_n,\omega) {+} {\rm H.a.}$.
 The correlation function  $C_{ijkl}^{nm}(\tau){=} \frac{1}{\hbar\pi\epsilon_0}\int\mathrm{d}\omega e^{-i\omega \tau}\mathbf{d}^{(n)}_{ij}\cdot\mathrm{Im}(G(\mathbf{r}_n,\mathbf{r}_m,\omega))   \cdot \mathbf{d}^{*(m)}_{kl}$ characterizes the system-bath coupling and dynamics and therefore is also present in the decomposed tensors $G$  \cite{supplemental} via $G_{ijkl}^{nm (\tilde{i})}{=}{-}\int_{t_{\tilde{i}}}^{t_{\tilde{i}+1}}\mathrm{d}\tau_1 \int_{t_{\tilde{i}}}^{\tau}\mathrm{d}\tau_1 e^{i((\epsilon_i-\epsilon_j)\tau_2+(\epsilon_k-\epsilon_l)\tau_1}C^{nm}_{ijkl}(\tau_2-\tau_1)$
 and 
 $G_{ijkl}^{nm (\tilde{i}\tilde{j})}{=}{-}\int_{t_{\tilde{i}}}^{t_{\tilde{i}{+}1}}\mathrm{d}\tau_1 \int_{t_{\tilde{j}}}^{t_{\tilde{j}{+}1}}\mathrm{d}\tau_1
e^{i(\epsilon_i{-}\epsilon_j)\tau_2+i(\epsilon_k{-}\epsilon_l)\tau_1)}$
$C^{nm}_{ijkl}(\tau_2-\tau_1)$.}

\red{For our example system shown in Fig.~\ref{schemesystem}(b), the TLSs 
are chosen resonant with the cavity modes.} 
\red{
We excite the TLSs by a few-\rm{fs} pulse described by $H_{\rm ext}=\sum_{l} \mathbf{d}\cdot \mathbf{E}(t) e^{i\omega_l t} |2\rangle_l \langle 1|_l+{\rm H.a.}$. The pulse does not excite the photons directly.
A Fourier transform of the TLS polarization is shown in Fig.~\ref{fig_qnmspec}(a).
}
\red{\footnote{The maximum bond dimension of the MPS in all numerical calculations is truncated to below 200 and an accuracy of $\epsilon=10^{-12}$, $\Delta T$ convergence is discussed in \cite{supplemental}.}}  
The low intensity plot (linear spectrum)
shows modified Rabi splitting between peaks 1 and 2 (cf.~Ref. \onlinecite{Yao:09}) and the upper state (2) shows a lifetime broadening due to a transfer process to a lower state (e.g. state $1$).  \red{ A longer delay (increasing $L$, with the same phase) between the TLS affects the upper state, caused by inter cavity transfer of photons. So the long inter cavity delay } turns the upper state to a sub radiant state with reduced broadening.
 Thus, the two cavities do not act as \red{a single} 
effective JCM model anymore;  this is also true for the upper state in the higher rungs of the JCM ladder, which we study below.

For nonlinear excitation, shown also in 
Fig.~\ref{fig_qnmspec}(a) ($0.45\pi$, half excited), we observe additional (negative) peaks 3 and 4. A comparison to the energies in the single and double quantum function, in Fig.~\ref{fig_qnmspec}(b),indicates that these match well with transitions between the first and second rung of the JCM ladder. The single and double quantum function give the coherences of a system between ground state and single or double excitation states  (respectively)~\cite{PhysRevLett.100.057402,kim2009two,PhysRevB.86.085308,Schlosser_2013} \red{and allows a direct inspection of their energies} 
(see  \cite{supplemental} 
\red{for more details on our quantum correlation functions}).
Other (positive) contributions between the first and second rung are overlapping with 
resonances 1 and 2, which slightly affect their line shape.
\red{The negative peaks 3 and 4 appear only in our nonlinear solution, showing complex interference effects beyond weak excitation. Although the origin of such line shapes are hard to identify, they may be related to an excitation transfer process~\cite{supplemental}.}
The time dynamics of densities in 
Fig.~\ref{fig_qnmspec}(d) shows the typical Rabi oscillations. 
Furthermore  correlations between the two TLS densities
(Fig.~\ref{fig_qnmspec}(d)), \red{allow access to complex entanglement properties}.

We highlight that for many  example systems, the bath correlation time is longer than the simulation time---a notoriously difficult  test for the numerical complexity  without  augmenting the density matrix \cite{vagov2011real}, yielding simulation times of days or weeks  depending on the bath correlation time and excitation.
Overall the example demonstrates the potential for eTEMPO algorithms where off-diagonal coupling is important to include.

In summary, we have provided a generalized version of the eTEMPO algorithm \cite{PhysRevLett.123.240602ETEMPO} to include off-diagonal system bath coupling, opening the route for numerically exact treatment of off-diagonal system bath coupling in exciton migration (e.g. coupled nanostructures, photosynthesis) or quantum light propagation in plasmonics and photonic networks. In particular, we have shown how delay and round trip coherence alters the JCM like behavior.

{\it Acknowledgments}---We  acknowledge funding from Queen's University,
the Canadian Foundation for Innovation, 
the Natural Sciences and Engineering Research Council of Canada, and support from 
the Alexander von Humboldt Foundation through a Humboldt Research Award.


\bibliography{pathintegral}


\clearpage
\begin{widetext}
\appendix
\setcounter{page}{1}
\setcounter{figure}{0}
\setcounter{equation}{0}
\renewcommand{\theequation}{S\arabic{equation}}
\renewcommand{\thefigure}{S\arabic{figure}}
\renewcommand{\thepage}{S\arabic{page}}
\renewcommand{\appendixname}{Supplemental material}

\section{Decomposition of tensors}

Here, we will discuss the decomposition of $\tilde{M}$ and $\tilde{E}$ into products of lower dimensional tensors and the underlying ideas.

We start with $\tilde{M}$ and its decomposition:
\begin{align} 
\tilde{M}_{p s k\alpha_{p}\alpha_{s}l}^{k' \alpha'_{p}\alpha'_{s}l'}(n)=&\sum_{\substack{\tilde{s}_1 \tilde{p}_1 \\ \tilde{p}_2 m_2 m_t}} \tilde{M}^{(1),\alpha'_{p}}_{\tilde{p}_2,\alpha_{p}m_t m_2} \tilde{M}^{(2) k' \tilde{p}_2 \alpha'_{s}l'm_t}_{\tilde{s}_1  \tilde{p}_1  k\alpha_{s} m_2}(n)
\tilde{M}^{(3)l'}_{m_2 m_t l} \tilde{M}^{(4)}_{\tilde{s}_1 s m_t}\tilde{M}^{(5)}_{\tilde{p}_1 p m_t}\tilde{M}^{(6)}_{s p m_t}. \label{m_decomp}
\end{align}
As seen in Eq.~\eqref{mtildetensor} the indices $k$ and $k'$ discriminate the different interactions handled in the tensor $\tilde{M}$. If $k$ is not changed ($\delta_{k,k'}$) $\tilde{M}$ does not change anything and just passes the information through present in the indices $\alpha_p$ $\alpha_s$, $k$ and $l$, in this case the interaction, if any,  happens in a subsequent tensor. If the index $k$ is changed from $1$ to $-1$ in $k'$, the double time integrated correlation function in $G$ describes a first system-bath interaction in the interval of the prior $\tilde{E}$ tensor, which is then finalized in the time interval connected to the $\tilde{M}$ tensor. Then the index $l$ passes the information, if the system-bath interaction happened on the left or right side of the Liouville space in $\tilde{E}$ to the tensor $\tilde{M}$, whereas $\alpha_s$ and $\alpha_p$ pass information about the changed system Hilbert space states described in $\tilde{E}$ to the tensor $\tilde{M}$. Therefore $\alpha'_s$ and $\alpha'_p$, which are passed to subsequent tensors $\tilde{M}$, are set to $1$ (note we use indices for s and p starting with $1$), which means the interaction already occurred and no subsequent interaction can occur on one of the remaining $\tilde{M}$.

The tensors used for the decomposition of $\tilde{M}$ reduce the number of indices per tensor and dimension of each of the indices.
We start with $\tilde{M}^{(1)}$:
\begin{align}
\tilde{M}^{(1),\alpha'_{p}}_{\tilde{p}_2,\alpha_{p}m_t m_2}=&(\delta_{\alpha_{p},\alpha'_{p}}\delta_{\tilde{p}_2,1}\delta_{m_t,3}\delta_{m_2,1}
+\sum_{m_1=1,2}\delta_{m_2,1}\delta_{m_t,m_1}\delta_{\tilde{p}_2+1,\alpha_{p}}\delta_{\alpha'_p,1}). 
\end{align}
 The value of the index $m_t$ determines 
whether $\tilde{M}$ does not act on the system ($m_t=3$) and thus just passes  $\alpha_{p}$ through to $\alpha'_{p}$, or whether  $\alpha_p$ is passed via the index $\tilde{p}_2$ to the integrated correlation function $G$ ($m_t=1,2$). $m_1$ indicates if the second interaction is on the left or right side and is passed to $m_t$.
In the next tensor $\tilde{M}^{(2)}$, we include the double integrated correlation function $G$:
\begin{align}
\tilde{M}^{(2),k',\tilde{p}_2,\alpha'_{s}l'm_t}_{\tilde{s}_1, \tilde{p}_1, k,\alpha_{s} m_2}(n)=&  \left(\delta_{\alpha_{s},\alpha'_{s}}\delta_{\tilde{s}_1,1}\delta_{\tilde{p}_1,1}\delta_{\tilde{p}_2,1}\delta_{m_2,1} \delta_{m_t,3}\delta_{k,k'}\right.
+\sum_{m_1=1,2}\sum_{\tilde{s}_2}\delta_{\tilde{s}_2+1,\alpha_{s}}\delta_{\alpha'_s,1} \delta_{m_t,m_1}\delta_{k',-1}\delta_{k,0}\nonumber\\ &\left.\qquad \cdot G_{\tilde{s}_1,\tilde{p}_1,\tilde{s}_2,\tilde{p}_2 m_1m_2}(n)\right). \label{G_tensor1}   
\end{align}
Again, for $m_t=3$ the incoming $\alpha_{s}$ is just passed through to $\alpha'_{s}$.
Whereas for $m_t=m_1=1/2$  a second system bath interaction on the left/right side in Liouville space in $G$ occurs. Therefore $\alpha_s$ is connected to $G$ via the index $\tilde{s}_2$.
$G$ is included in $\tilde{M}^{(2)}$ and its other indices are passed to the other tensors in the product: $\tilde{p}_2$ to $\tilde{M}^{(1)}$, $\tilde{p}_1$ to $\tilde{M}^{(5)}$, $\tilde{s}_1$ to $\tilde{M}^{(4)}$, $m_2$ to $\tilde{M}^{(3)}$.
The next tensor $\tilde{M}^{(3)}$ handles the distribution of the $l$ index:
\begin{align}
\tilde{M}^{(3)l'}_{m_2 m_t l}=&\delta_{m2,1}\delta_{l,l'}\delta_{m_t,3} +\sum_{m_1=1,2} \delta_{m_1,m_t}\delta_{m_2+1,l}\delta_{l',1}.
\end{align}

As usual, for $m_t=3$, $l$ is passed through to $l'$, and for $m_t=m_1=1,2$  $l$ is passed to  $G$.
The next tensors starting with $\tilde{M}^{(4)}$ connect indices coming from $G$ either with the left or right part of the indices $s$, $p$: 
\begin{align}
\tilde{M}^{(4)}_{\tilde{s}_1 s m_t}=& \delta_{m_t,1} \delta_{s_L,\tilde{s}_1}-\delta_{m_t,2} \delta_{s_R,\tilde{s}_1}+\delta_{m_t,3}\delta_{\tilde{s}_1,1},  
\end{align}
so for $m_t=1$ $\tilde{s}_1$ is connected to $s_L$ and for $m_t=2$ $\tilde{s}_1$ is connected to $s_R$. $m_t=3$ is the pass-through mode and effectively does nothing.
$\tilde{M}^{(5)}$ is constructed in an analogous way:
\begin{align}
\tilde{M}^{(5)}_{\tilde{p}_1 p m_t}=& \delta_{m_t,1} \delta_{p_L,\tilde{p}_1}-\delta_{m_t,2} \delta_{p_R,\tilde{p}_1}+\delta_{m_t,3}\delta_{\tilde{p}_1,1}, 
\end{align}
since for $m_t=1$ $\tilde{p}_1$ is connected to $p_L$ and  for $m_t=2$ $\tilde{p}_1$   to $p_R$.
The next tensor $\tilde{M}^{(6)}$  connects  $s$ and $p$ either to the left or right part (on opposite site to the previous tensors!), and is controlled again by $m_t$:
\begin{align}
\tilde{M}^{(6)}_{s p m_t}=&\delta_{m_t,1}\delta_{s_R,p_R}+\delta_{m_t,2}\delta_{s_L,p_L}+\delta_{m_t,3}. 
\end{align} 
Thus for $m_t=3$ nothing happens, and for $m_t=1$ the right sides of $s$ and $p$ are connected, whereas for $m_t=2$ the left sides are connected.

Before we discuss the decomposition of $\tilde{E}$, we first discuss its contributions from Eq.(\ref{etildetensor}):
For $k=0$ no interaction is happening in the interval $\Delta T$, so we do not have to pass information to the $\tilde{M}$ tensors.
On the other hand $k=-1$ means, so far nothing is applied to the time interval, the contribution will occur in one of the $\tilde{M}$, which are later added to interval $\Delta T$, therefore $k=-1$ will not be included in the summation at the edge of network.
Also $k=2$ means no system bath interactions in subsequent $\tilde{M}$ inside $\Delta T$, which are connected to other $\Delta T$ intervals.
Finally for $k=1$  one system-bath interactions occur in the $\Delta T$ interval associated with $\tilde{E}$, whereas the second occurs in another $\Delta T$ interval involving its $\tilde{M}$ tensor. Here $l$ encodes the side of the Liouville space (left or right), on which the system bath interaction occurs.

Again the tensors used for the decomposition of $\tilde{E}$ into low rank and dimensional tensors read:
\begin{align} 
&\tilde{E}_{s,p,k}^{\alpha_{p},\alpha_{s}l }= \sum_{m_1 m_2 q_1 q_2 s_1 p_2} E^{(1)}_{p_Ls_Lkm_1m_2}E^{(2)}_{p_Rs_Rkm_1m_2}E^{(3)}_{p_Lp_R q_1 q_2}
E^{(4)}_{s_Ls_R m_1 m_2  q_2} E^{(5)}_{m_1 m_2 s_1 q_1 q_2 p_2 l} E^{(6)}_{s_Ls_R\alpha_s s_1 m_1 m_2} E^{(7)}_{p_Lp_R\alpha_p p_2 m_1 m_2}, \label{etildecomp}
\end{align}
 which reduces the number and dimension of the used indices, and thus made the numerical calculations more feasible.

The first tensor, $E^{(1)}$, in the product
\begin{align}
E^{(1)}_{p_Ls_Lkm_1m_2}=&\delta_{k,0}\delta_{p_L,s_L}\delta_{m_1,1}\delta_{m_2,1}+\delta_{k,-1}
+\delta_{k,2}(1-\delta_{m_1,2}\delta_{m_2,2})  
+\delta_{k,2}\delta_{m_1,2}  \delta_{m_2,2} \delta_{p_L,s_L}
+\delta_{k,1} (\delta_{m_2,1}-\delta_{m_2,2}\delta_{p_L,s_L}), 
\end{align}
connects  the left part $p_L$ and $s_L$,  if no system-bath interaction in $\Delta T$ occurs ($k=0$), if two interactions at the left side occur ($k=2$ with $m_1=2$ and $m_2=2$) or if one interaction on the right side takes place ($k=1$ with $m_2=1$).
The second tensor $E^{(2)}$:
\begin{align}
E^{(2)}_{p_Rs_Rkm_1m_2}=&\delta_{k,0}\delta_{p_R,s_R}\delta_{m_1,1}\delta_{m_2,1}+\delta_{k,-1} 
+\delta_{k,2}(1-\delta_{m_1,1}\delta_{m_2,1}) 
&+\delta_{k,2}\delta_{m_1,1}  \delta_{m_2,1} \delta_{p_R,s_R} 
+\delta_{k,1} (\delta_{m_2,1}\delta_{p_R,s_R}+\delta_{m_2,2}) 
\end{align}
works in the same way, only that it acts on the right side instead of the left side, thus right and left (also encoded in $1$ and $2$ for $m_1$ and $m_2$) is exchanged and a sign difference caused by the commutator.
The term,
$E^{(3)}$, is only important for two system bath interaction per $\Delta T$
\begin{align}
E^{(3)}_{p_Lp_R q_1 q_2}=& (\delta_{k,0}+\delta_{k,-1}+\delta_{k,1}) \delta_{q_1,1}\delta_{q_2,1} 
+\delta_{k,2}(\delta_{m_1m_2}\delta_{q_1,q_2} 
+\delta_{m_1,1}\delta_{m_2,2}\delta_{q_1,p_L} 
+\delta_{m_1,2}\delta_{m_2,1}\delta_{q_1,p_R}),
\end{align}
and only for $k=2$ is something  done.
It either connects the indices $q_1$ $q_2$ of the correlation function $G$, if the first and second system-bath interaction occurs both on the left or on the right side, or $q_1$ is connected either to the left or right side of $p$, these contributions are a result of the double commutator of the system bath coupling.
The next factor $E^{(4)}$ completes the work for the correlation function $G$:
\begin{align}
E^{(4)}_{s_Ls_R m_1 m_2  q_2}=& \big(\delta_{k,0}+\delta_{k,-1}+\delta_{k,1})\delta_{q_2,1}\delta_{m_1,1}\delta_{m_2,1}
-\delta_{m_1,2}\delta_{m_2,1} \delta_{q_2,s_L} +\delta_{m_1,m_2} )  
\end{align}
and connects for $k=2$ with interaction on left and right the index $q_2$ either to the left or right side of $s$.
The next factor $E^{(5)}$ serves two purposes:
\begin{align}
E^{(5)}_{m_1 m_2 s_1 q_1 q_2 p_2 l}=&  ((\delta_{k,0}+\delta_{k,-1})\delta_{m_2,1} \delta_{l,1} +\delta_{k,1}\delta_{m_2,l+1}) 
-\delta_{k,2}  \delta_{l,1} G_{s_1,q_1,q_2,p_2,m_1,m_2}, \label{G_tensor2} 
\end{align}
which contains the double integrated correlation function $G$ for the two interaction in $\Delta T$ (case $k=2$) and for $k=1$ it connects $l$ shifted by one to $m_2$ to pass it on to the $M$ tensors.
The indices $q_1$ and $q_2$ are passed to $G$ from the previous tensors, where as $s_1$ and $p_2$ will come from the proceeding tensors, starting with $E^{(6)}$:
\begin{align}
E^{(6)}_{s_Ls_R\alpha_s s_1 m_1 m_2}=& (\delta_{k,0}+ \delta_{k,-1} ) \delta_{\alpha_s,1} \delta_{m_1,1}\delta_{m_2,1}\delta_{s_1,1}
+\delta_{k,2} \delta_{\alpha_s,1} (\delta_{m_1,1}\delta_{s_1,s_L}+\delta_{m_1,2}\delta_{s_1,s_R} ) 
\nonumber\\ &
+\delta_{k,1}\delta_{m_1,1}(\delta_{m_2,1}\delta_{s_L,\alpha_s+1}\delta_{s_1,1}  
+\delta_{m_2,2}\delta_{s_R,\alpha_s+1}\delta_{s_1,1}), 
\end{align}
where for $k=2$, $s_1$ is depending on $m_1$ either to the left or right side of $s$.
For $k=1$ the left or right side of $s$ (depending on $m_2$) is connected to $\alpha_s$ shifted by $1$, to connect it to a $\tilde{M}$ tensor for the second system-bath interaction.
The last factor $E^{(7)}$ works in a similar way:
\begin{align}
E^{(7)}_{p_Lp_R\alpha_p p_2 m_1 m_2}=&(\delta_{k,0}+ \delta_{k,-1} ) \delta_{\alpha_s,1} \delta_{m_1,1}\delta_{m_2,1}\delta_{s_1,1} 
+\delta_{k,2} \delta_{\alpha_p,1} (\delta_{m_2,1}\delta_{p_2,p_L}+\delta_{m_2,2}\delta_{p_2,p_R} ) 
\nonumber\\ &
+\delta_{k,1}\delta_{m_1,1}(\delta_{m_2,1}\delta_{p_L,\alpha_p+1}\delta_{p_2,1} 
+\delta_{m_2,2}\delta_{p_R,\alpha_p+1}\delta_{p_2,1}), 
\end{align}
for $k=2$ it connects either the left or right side (depending on $m_2$) of $p$ to $p_2$ for passing it to the correlation function $g$. For $k=1$ the left or right side of $p$ is passed  to $\alpha_p$ shifted by $1$. 

\section{Green functions and model for the photonic crystal cavity system}

The electromagnetic response functions can be defined in terms of the photonic Green functions, defined from
\begin{equation}
\boldsymbol{\nabla}\times\boldsymbol{\nabla}\times\mathbf{G}(\mathbf{r},\mathbf{r}',\omega) -\frac{\omega^2}{c^2}\epsilon(\mathbf{r},\omega)\mathbf{G}(\mathbf{r},\mathbf{r}',\omega)=\mathbb{1}\delta(\mathbf{r}-\mathbf{r}')\label{eq: GFHelmholtz2},
\end{equation}
with suitable boundary conditions.
Following the approach of
Refs.~\onlinecite{Yao:09,PhysRevLett.98.083603}, these can be obtained
in an analytical form using mode expansion techniques and exploiting the completeness relation of the Green function, and we consider the two cavity modes and the waveguides modes as the only modes of interest in the problem. We assume identical cavities, that are side-coupled to a waveguide mode, and separated by a distance $L$;
we also consider a TLS embedded in each cavity at a field antinode position, aligned with the polarization of the
cavity mode. Subsequently, the 
 correlation functions,
 defined from
 $G_{nm}(\omega)=\mathbf{d}^{(n)}_{12}{\cdot}\mathrm{Im}({\bf G}(\mathbf{r}_1,\mathbf{r}_1,\omega))   \cdot \mathbf{d}^{*(m)}_{12}$, 
 take the form \cite{Yao:09} 
\begin{align}
    &G_{11}(\omega)=\frac{\omega^2 E_1 E_1}{\omega_{c}^2-\omega^2-i \omega\gamma_{c}(1+e^{i2\phi}R_2(\omega))-i\omega\gamma_0}, \\
    &G_{22}(\omega)=\frac{\omega^2 E_2 E_2}{\omega_{c}^2-\omega^2-i \omega\gamma_{c}(1+e^{i2\phi}R_1(\omega))-i\omega\gamma_0}, \\
    &G_{12}(\omega) = \frac{\omega^2R_1(\omega) e^{i\phi}E_1 E_2}{\omega_c^2-\omega^2-i \omega \gamma_c(1+e^{i2\phi}R_1(\omega))-i \omega \gamma_0}, \\
    &G_{21}(\omega) = \frac{\omega^2R_2(\omega) e^{i\phi}E_1 E_2}{\omega_c^2-\omega^2-i \omega \gamma_c(1+e^{i2\phi}R_2(\omega))-i \omega \gamma_0}
\end{align}
with $E_{1/2}=\frac1{\epsilon_rV_{\rm eff}}$ (with $V_{\rm eff}$  the effective mode volume and $\epsilon_r$  the dielectric constant of the photonic crystal slab), $R_{1}(\omega)= R_2(\omega) = \frac{i\omega\gamma_c}{\omega_c^2-\omega^2-i\omega(\gamma_c+\gamma_0)}$, and $\gamma_c=Q_c/\omega_c$. The cavity decay rates include
on-chip decay from the cavity to the waveguide, $\gamma_c$, as well as off-chip
decay, $\gamma_0$, where the latter is typically much smaller than the former (to maintain a waveguide mode beta factor). The round trip phase
is determined from
$2\phi = 2k_BL + \varphi$, $k_B=\omega n_g/c$
and we set $\varphi=0.8 \pi$.

Parameters used in  the main text (if not others specified): 
$\epsilon_r=12$, $L=2 \lambda$ (waveguide length in wavelengths)
$n_g=10$ (group velocity index), 
$\omega_c=200\mathrm{THz}$ frequency of the cavities, $Q_c=2000$ quality factor of the cavity, $V_{\rm eff}=5\cdot 10^{7} {\rm nm}^3$ mode volume, $\varphi=0.8 \pi$ 
out of plane loss $1\%$ of cavity loss ($\gamma_0=0.01\gamma_c$).

\section{Dipole correlation function and localized detection}
\begin{figure}[bt]
	\includegraphics[width=0.6\textwidth]{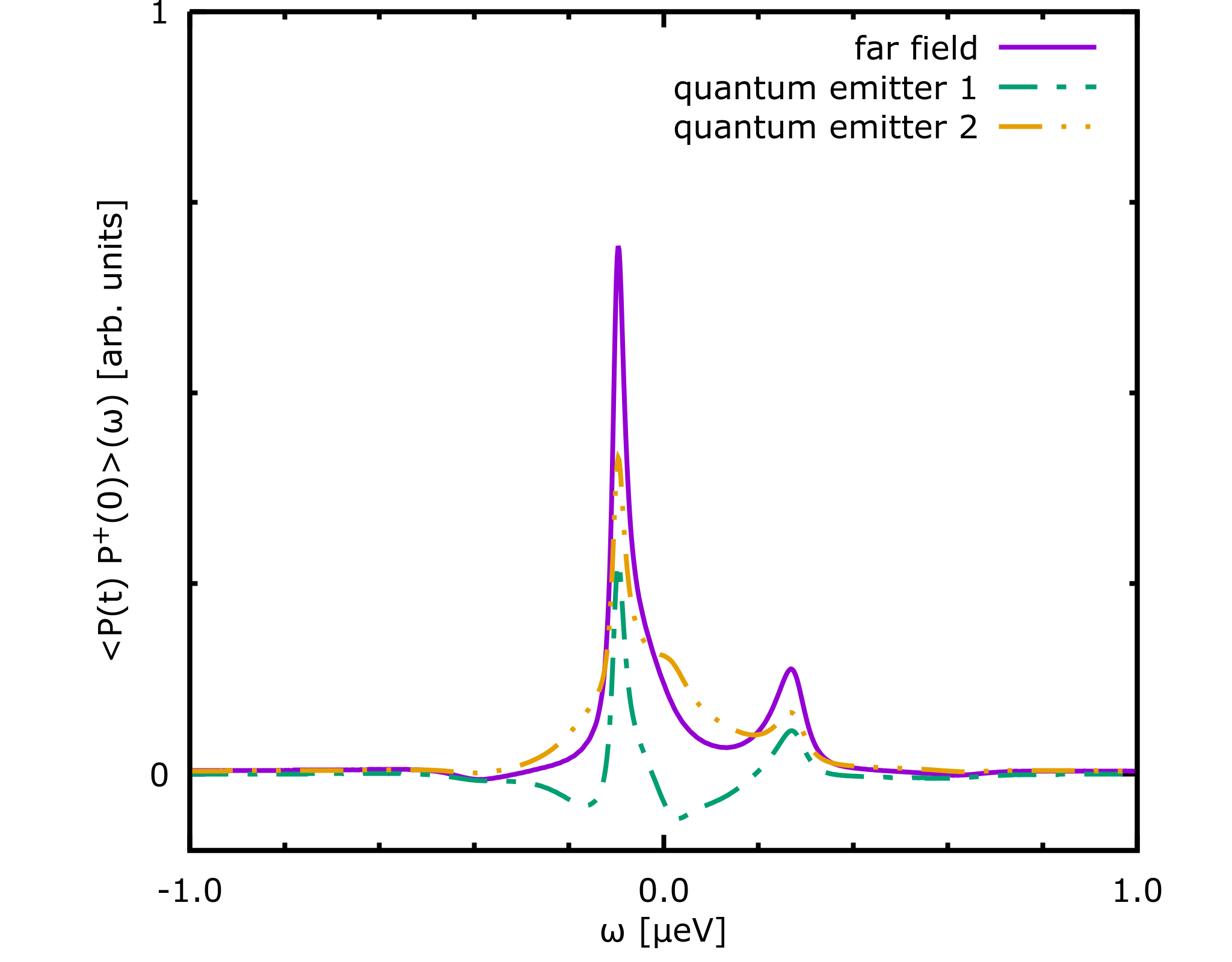}
	\caption{Dipole correlation function $\langle P(t)  P^{\dagger}(0)\rangle $ for a system initially on an excited density at emitter 1 for far field detection (detection in $P(t)$ contains both emitters) and for localized detection at the quantum emitter (detection in $P(t)$ contains only one emitter).  }

	\label{fig_qnmspec_sub}
\end{figure}
The single quantum 
$\langle P(t)  P^{\dagger}(0)\rangle $ and double quantum  dipole correlation function $\langle P(t) P(t)P^{\dagger}(0) P^{\dagger}(0)\rangle$ allow a direct inspection of the system eigenenergies, and spectral resonances, for one or two excitations in the system, if the system is initially in the ground state. For example, $\langle P(t)  P^{\dagger}(0)\rangle $ describes a coherence between the ground state and single excited states created by polarisation flip operator $P^{\dagger}(0)$ at time $0$ and inspected at time $t$ by $P(t)$. Thus by varying $t$, the coherent oscillation can be scanned, and its Fourier transform corresponds to the linear optical response and we can deduce from the individual resonances the bright single excited states of system.

On the other hand, for $\langle P(t) P(t)P^{\dagger}(0) P^{\dagger}(0)\rangle$ at time $0$,  $P^{\dagger}(0) P^{\dagger}(0)$ described the coherence between the ground state and double excited states; similar to the linear, single excitation case, the double excited states can be scanned and a Fourier transform retrieves the energies of the  double excited states available for two photon excitation.
As in the linear case, experimental techniques exist to access this correlation function \cite{kim2009two}.

However, for the example system
in the main Letter,
the situation is a bit different to the usual experimental situation, since we include only polarization operators of the quantum emitters, for the operators $P^\dagger$. But in the system, the coupling to the photons can be  strong, so that eigenstates are in fact polariton states, that also include a photon mode part.
Assuming photon mode operators $b^\dagger_j$, then the single excitation eigenstates $|e\rangle$ are a superposition state,
\begin{align}
    |e\rangle=c^1_e \sigma_1^\dagger |0\rangle +c^2_e \sigma_2^\dagger |0\rangle+ \sum_j c^{pj}_e b^\dagger_j , 
\end{align}
with $\sigma_l^\dagger= |2\rangle_l \langle 1|_l$ and an overall ground state $|0\rangle$. So it is clear that we miss the photon part in the spectra.
In an analogues way, the double excited eigenstates $|f\rangle$ can be expanded in the local states.

In an experimental pump probe experiment, the pulses are also only acting on the quantum emitters and are much faster compared to the transfer process to the photon modes, thus that local densities $|2\rangle_l \langle 2|_l$ are prepared in the excited system.
We can access this contribution also by inspecting $\langle P(t)  P^{\dagger}(0)\rangle $ for a system where the first quantum emitter is initially in the local excited state. Of course this is not an eigenstate of the system but a superposition of $|e\rangle$, furthermore after the polarisation operator at $t$ the system is in a coherence between states  $|f\rangle$ and $|e\rangle$, creating the second excitation in the other emitter.
We see in Fig.~\ref{fig_qnmspec_sub}  precisely the same features as in the nonlinear signal in the main text.  The negative features are of course puzzling, but can be a result, that the system is in the local basis and thus expansion coefficients such as $c^1_e$ or $c^{12}_f$ etc, and their phase enter the signal. While our method does not allow to access the bath---the photon part---we can show, that if we detect only one quantum emitter \cite{localized_excitation,PhysRevB.86.085308,Schlosser_2013,specht2016reconstruction}, formally dark state can give negative contributions in the spectrum (cf.~Fig.~\ref{fig_qnmspec_sub}) and likely this is also the case for the negative contributions with respect to the photon part.

\section{Convergence analysis and consequences of approximate unitary}

\begin{figure}[hbt]
	\includegraphics[width=0.6\textwidth]{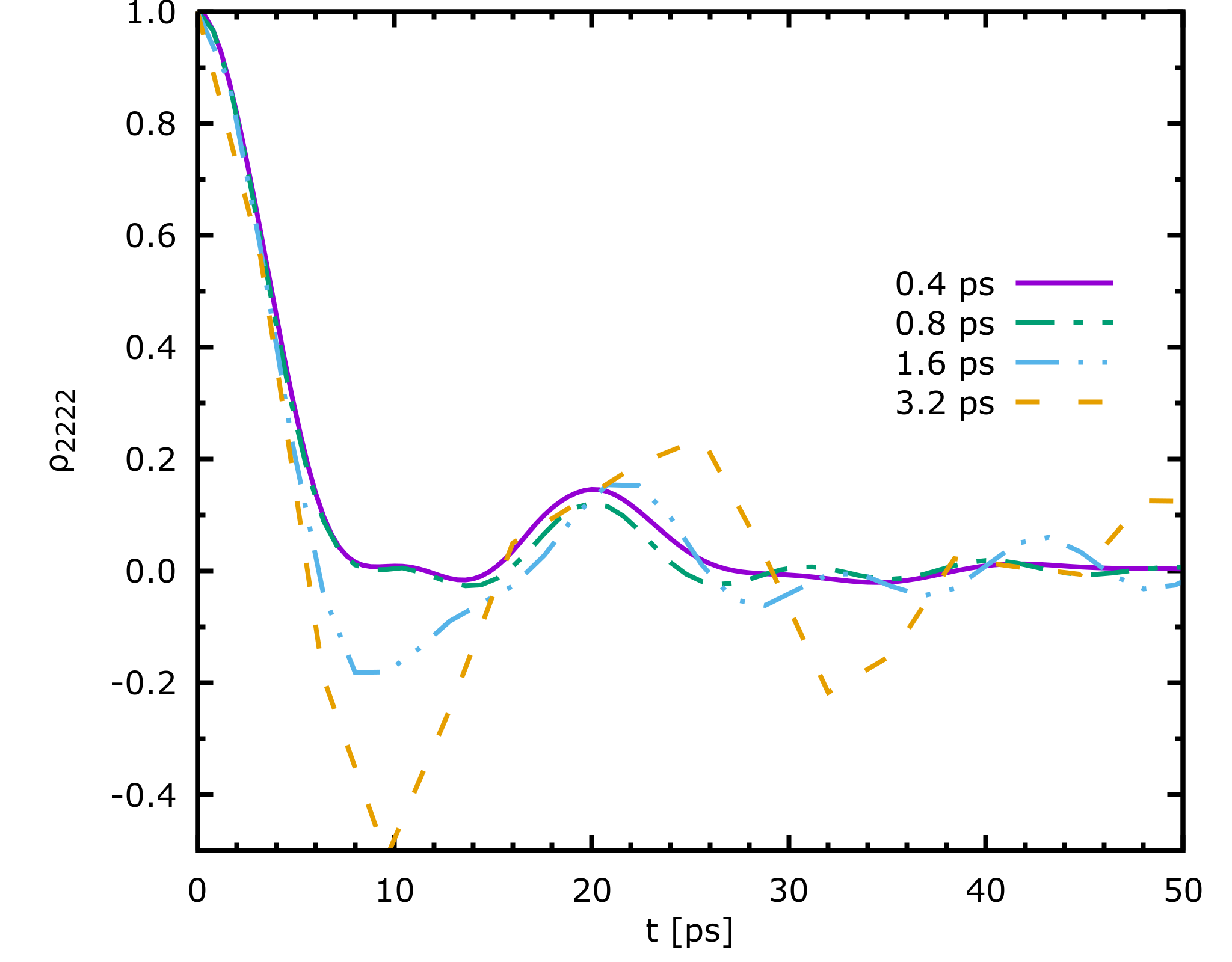}
	\caption{Reduced density matrix element $\rho_{2222}$ for the example from Fig. \ref{fig_qnmspec} d) for different discretizations $\Delta t$ of the quantum path integral. }

	\label{fig_qnmspec_sub_conv}
\end{figure}
Main parameters for the numerically convergence are the truncation bond dimensions and accuracy for the tensor network and the quantum path integral time discretization $\Delta T$.

For truncating bonds the MPS after MPO application, we use a maximum bond dimension as well as a accuracy for determining the truncated dimension. This is necessary since the different layers of the network within a time step have quite different bond dimensions (one order magnitude possible).
So for some of them bond dimension is the best parameter to do the truncation and for the others accuracy is the best for keeping the important information in the network. This is at least our subjective experience while designing the algorithm. Some intermediate layers require relative low bond dimension.  There relative useless information for the next will be included without an accuracy truncation parameter. This will prevent a fast calculation in subsequent layers, which require higher bond dimensions cause by the additional useless information.
Even worse, it could prevent convergence at all.

The time propagation  $U_i$ over an interval $\Delta T$ is expanded in Eq. \eqref{uexpansion} in orders of the system-bath interaction, thus it is only $O(\Delta t^2)$ unitary. In principle this can have consequences on semi-positivity and trace preservation of the reduced density matrix thus the density matrix elements over time as plotted in Fig. \ref{fig_qnmspec} d).

In fact, if we choose a $\Delta T$, which is too large,
we see large negative values e.g. for the reduced density of the double excited state matrix $\rho_{2222}$ as shown in Fig. \ref{fig_qnmspec_sub_conv}, but these problems disappear for sufficiently small $\Delta T$.
Problems with violation of density matrix positivity also occur if too strong truncation removes two many quantum path ways.
The violated positivity in $\rho_{2222}$ can also be seen in the numerical calculation used for Fig. \ref{fig_qnmspec}(a), but it occurs after the polarizations are created, so it can not be the origin of the negative signatures, since only secular terms are included in the system bath coupling, which do not convert polarizations into densities and vice versa. Therefore the densities are completely decoupled from the plotted observables.

\end{widetext}


\end{document}